# Emission of Circularly Polarized Terahertz Wave from Inhomogeneous Intrinsic Josephson Junctions

Hidehiro Asai and Shiro Kawabata

*Abstract*—We have theoretically demonstrated the emission of circularly-polarized terahertz (THz) waves from intrinsic Josephson junctions (IJJs) which is locally heated by an external heat source such as the laser irradiation. We focus on a mesa-structured IJJ whose geometry is slightly deviate from a square and find that the local heating make it possible to emit circularly-polarized THz waves. In this mesa, the inhomogeneity of critical current density induced by the local heating excites the electromagnetic cavity modes TM (1,0) and TM (0,1), whose polarizations are orthogonal to each other. The mixture of these modes results in the generation of circularly-polarized THz waves. We also show that the circular polarization dramatically changes with the applied voltage. The emitter based on IJJs can emit circularly-polarized and continuum THz waves by the local heating, and will be useful for various technological application.

## I. Introduction

Terahertz (THz) electromagnetic (EM) wave has huge potential to open up novel technological applications such as the non destructive inspection of materials, medical diagnosis, bio sensing, and high-speed wireless-communication [1]. Hence, in recent years, the generation of intense THz waves has been a hot topic for such applications, and various kinds of THz emitter have been extensively studied [2], [3].

Recently, a single crystal of $Bi_2Sr_2CaCu_2O_{8+x}$ (Bi2212) has attracted much attention as a one of promising candidate for compact solid-state THz-source. Superconducting $CuO_2$ layers and insulating Bi-Sr-O layers in the Bi2212 crystal naturally form a stack of Josephson junctions so called intrinsic Josephson junctions (IJJs). A voltage applied to the Bi2212 crystal can generate AC Josephson current in a THz frequency range, and this AC current gives rise to the THz wave emission. Since intense THz emission has been reported for Bi2212 fabricated in a mesa geometry, a number of studies have focused on THz emission from mesa-structured IJJs [4]-

This work was partially supported by Grant-in-Aid for Young Scientists (B) from JSPS (Grant No. 34026790062) and a Grant-in-Aid for Scientific Research (C) from JSPS (Grant No. 24510146).

Hidehiro Asai is with the National Institute of Advenced Industrial Science and Technology, Tsukuba city, Ibaraki prefecture, 305-8568 Japan (e-mail: hd-asai@aist.go.jp).
Shiro Kawabata is with the National Institute of Advenced Science and Technology, Tsukuba city, Ibaraki prefecture, 305-8568 Japan (e-mail: hd-asai@aist.go.jp).

[30]. The mesa structure whose dimensions are 50~500 μm behaves as a cavity resonator for THz waves and amplify the THz emission power from the IJJs. The observed emission power has reached the order of 30 μW for the single IJJ mesa [13] at this stage, and has continued to rise. However, the polarization properties of the THz waves from IJJs are still poorly understood. Controlling the polarization of THz waves is a one of key issue in various applications of THz waves. In particular, generation and control of circularly-polarized THz wave is crucial for the wireless communication as well as the bio-medical application.

In this paper, we theoretically investigate the emission of circularly-polarized THz waves from a IJJ mesa whose geometry is slightly deviate from a square. We assume the IJJ mesa has inhomogeneous distribution of critical current density $j_c$ corresponding to the laser heating. In our previous theoretical study, we pointed out that the inhomogeneous $j_c$ distribution induced by laser heating strongly excites the Josephson plasma wave inside IJJs and amplify the THz emission power [29]. The control of $j_c$ distribution by laser heating has been realized in conventional Josephson junctions [31], [32], and recent experimental studies verify the laser heating amplifies the THz emission from the IJJ mesa [15], [16]. Thus, the laser heating is expected to be a powerful tool for controlling the THz emission from the IJJ mesa. In this study, we show the inhomogeneity of the $j_c$ at the corner of the mesa excite both the TM (1,0) and TM (0,1) cavity mode. Since the mesa is almost square, both the cavity modes, whose polarizations are orthogonal to each other, are mixed around the cavity resonant frequencies. We find the mixture of the cavity modes leads to the emission of circularly-polarized THz waves. We also demonstrate that the polarization remarkably changes with the applied voltage.

## II. Theory

Figure 1 shows the schematics of the IJJ mesa locally heated by laser irradiation. The Figs. 1(a) and (b) show projection views in the *x-z* plane and the *x-y* plane, respectively. The mesa is attached to an infinite ground plane and covered by a metal electrode. The upper electrode is heated by the focused laser beam. The temperature of the mesa beneath the electrode locally increases and the decrease the $j_c$ [29]. As a result, the averaged temperature of the mesa also increases by the laser heating. In this study, for simplicity, we simulate laser heating by decreasing the $j_c$ in a square shape

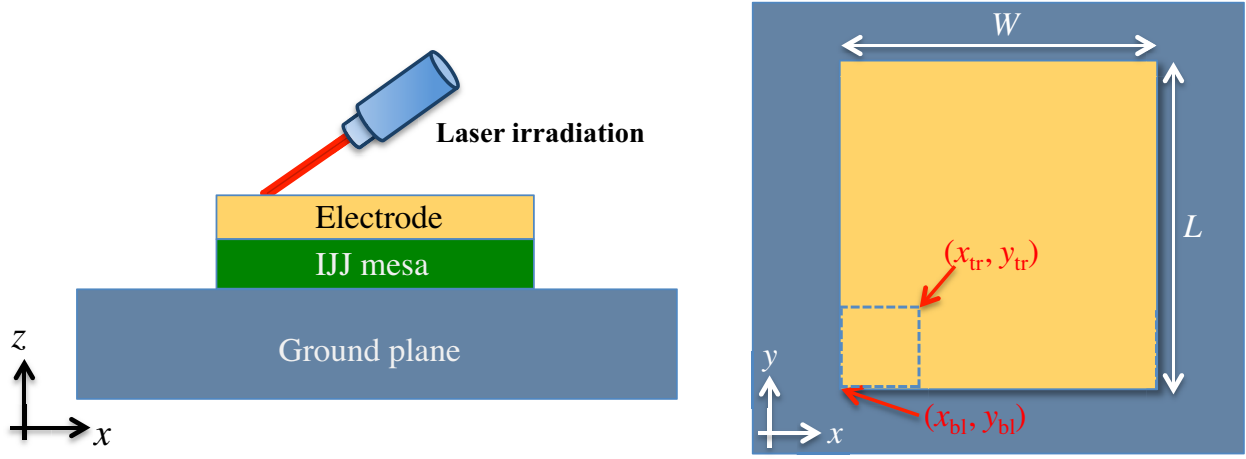

Fig. 1. Schematic figures of IJJ mesa whose geometry is almost square, (a) in x-z plane, and (b) in x-y plane. The dotted square in (b) indicates the position of the local heating defined by ($x_{tr}$, $y_{tr}$) and ($x_{bl}$, $y_{bl}$).

[26]. A dotted square in Fig. 1(b) indicates the heating position, and the position is identified by ($x_{tr}$, $y_{tr}$) and ($x_{bl}$, $y_{bl}$).

Here, we consider the mesa whose geometry is slightly deviate from a square. In such mesa, resonant frequencies of the cavity modes along the $x$ axis and the $y$ axis, i.e. TM (1,0) and TM (0,1) modes, are close to each other, and thus, these two modes can mix around the resonant frequencies. Moreover, the difference of the resonant frequencies of the cavity modes in such a mesa structure leads to the relative phase difference between TM (1,0) and TM (0,1) modes. The mixture of these modes with some phase difference results in the generation of the circularly or elliptically polarized THz waves. In this study, according to the classical antenna theory for circularly polarized waves [33], we take the dimensions of the mesa as follows for the typical c-axis conductivity of the IJJ, $\sigma_c$ = 10 S/m: the width $W$ and the length $L$ of the electrode and the IJJ mesa are $W$ = 78 μm, $L$ = 80 μm, the thickness of the electrode is 2 μm, and the thickness of the IJJ mesa is 2 μm.

Since the TM (1,0) and TM (0,1) modes can be strongly excited by the anti-symmetric $j_c$ distribution with respect to the $x$ axis and $y$ axis respectively [26], the local heating at the corner of the mesa as shown in Fig. 1 (b) is expected to cause the mixture of the cavity modes. Thus, we take the heating position ($x_{tr}$ = -29 μm, $y_{tr}$ = -30 μm) and ($x_{bl}$ = -39 μm, $y_{bl}$ = -40 μm ).

For examining THz emission from the IJJ mesa, we calculate dynamics of the superconducting phase inside of the mesa and the electromagnetic field outside of the mesa simultaneously. In this study, we use the in-phase approximation in which all phase differences are equal to the common phase difference $\phi$ [19], [22], [26] and the equation for phase difference $\phi$ becomes

$$\frac{\hbar \varepsilon_c}{2eD} \frac{\partial^2 \phi}{\partial t^2} = c^2 \left[ \frac{\partial B_y}{\partial x} - \frac{\partial B_x}{\partial y} \right] - \frac{1}{\varepsilon_0} [j_c(x,y) \sin\phi + \sigma_c E_z - j_{ex}], \quad (1)$$

where $D$ is the thickness of insulating layers of IJJ, $\varepsilon_c$ is the dielectric constant of the junctions, $c$ is the light velocity, $j_{ex}$ is the external current density, $j_c(x,y)$ is the critical current density. In this study, we take $j_c(x,y) = 0.5\, j_{c0}$ in the laser heating spot ($x_{bl} < x < x_{tr}$, $y_{bl} < y < y_{tr}$) and $j_c(x,y) = j_{c0}$ in the other region. The EM fields in the IJJs are given by

$$E_z = \frac{\hbar}{2eD} \frac{\partial \phi}{\partial t}, \quad (2a)$$

$$B_x = -\frac{\hbar}{2eD} \frac{\partial \phi}{\partial y}, \quad (2b)$$

$$B_y = \frac{\hbar}{2eD} \frac{\partial \phi}{\partial x}. \quad (2c)$$

In this calculation, we take $\varepsilon_c$ = 17.64, $D$ = 1.2 nm, and $j_{c0}$ = 2.18 · $10^3$ A/cm$^2$. The far fields are calculated from the equivalent electric and magnetic current along the surface of the calculation region [33].

### III. RESULTS AND DISCUSSION

Figure 2 (a) shows emission power as a function of voltage per IJJ layer for the mesa. The solid curve indicates the fitting curve of calculated emission power with two Lorentzian functions, and the dotted and the dashed curves indicate the each of the Lorentzian peak. We can see that two emission peaks appear around 0.923 mV and 0.941mV which correspond to the excitation of the TM (0,1) mode and the TM (1,0) mode, respectively. The peak emission powers are ~10 µW and comparable to those of the experimental studies [11], [13]. Of particular note is that the two peaks largely overlap as shown in Fig. 2(a). This is because the resonant frequencies of the cavity modes are close to each other. Thus, the emission power is high ~10 µW even at the middle of the two peak voltages 0.932 mV, in which the mixture of the TM (0,1) and the TM (1,0) mode takes place. Figure 2 (b) indicates the frequency spectrums of the emission waves at 0.923 mV, 0.932 mV, and 0.941 mV, respectively. As can be seen from this figure, the IJJ mesa emits EM waves in THz frequency range according to the AC Josephson relation $f_J = 2eV/h$. Moreover, the peak frequencies almost satisfy the cavity resonance condition $f_J = c/(2\sqrt{\varepsilon_c}L) = 0.446$ THz for the TM(0,1) mode and $f_J = c/(2\sqrt{\varepsilon_c}W) = 0.458$ THz for the TM(1,0) mode.

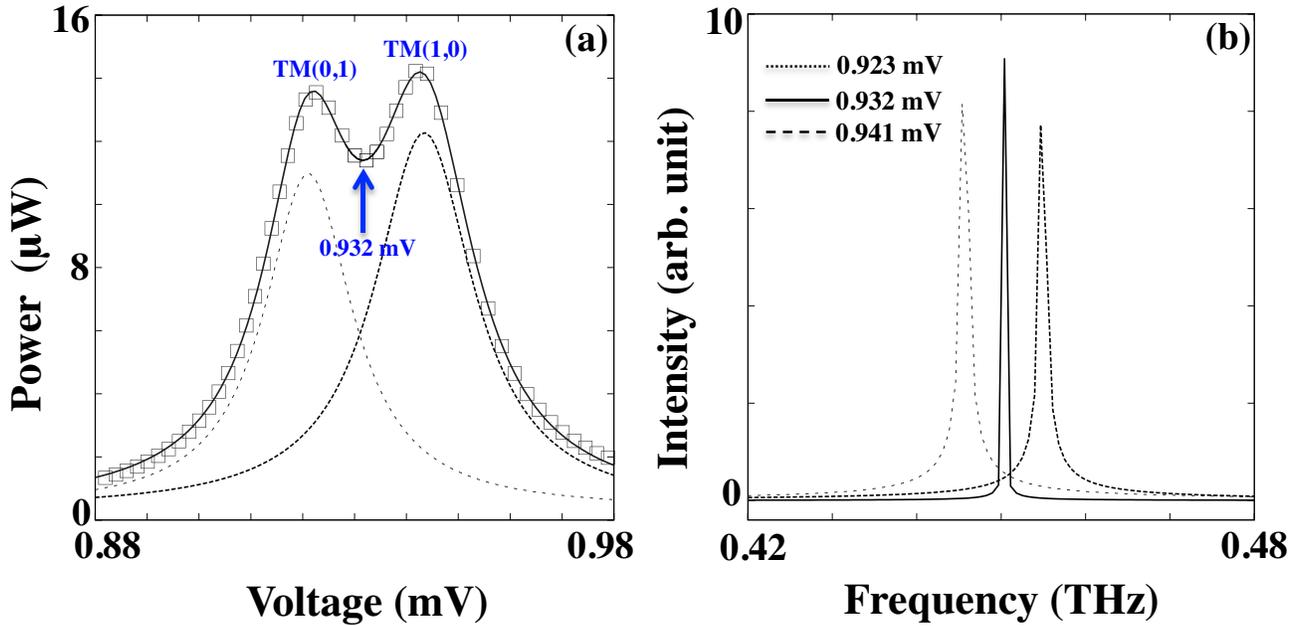

Fig. 2. (a) Emission power as a function of voltage. (b) Frequency spectrum of the electromagnetic wave emitted from the IJJ mesa.

Finally, in Figs. 3(a)-(c), we show the trajectories of far electric fields along the $z$ axis at 0.923 mV, 0.932 mV, and 0.941 mV. The arrows in the figures exhibit the direction of the time evolution of the fields. Figure 3(b) indicates that the circularly-polarized THz wave is observed at 0.932 mV because both the TM (0,1) and TM (1,0) mode mix each other. On the other hand, the elliptically-polarized THz waves are observed at 0.923 mV and 0.941 mV. Since the resonance frequencies of the TM (0,1) and TM (1,0) modes are almost identical, both the cavity modes are excited at these voltages. However, for example, the TM(0,1) mode is strongly excited compared to the TM (1,0) mode at 0.923 mV. Thus, the amplitude of the $y$ component becomes larger than that of x component as shown in Fig. 3(a). Our results strongly indicate that the polarization of the THz waves can be controllable by the applied voltage.

## IV. CONCLUSION

In summary, we have theoretically studied the polarization of terahertz (THz) waves emitted from a mesa-structured intrinsic Josephson Junction (IJJ) whose geometry is almost square. We have assumed that the corner of the mesa is locally heated by the laser irradiation, and found that circularly-polarized THz waves are emitted from the mesa. In this mesa, both the TM (0,1) and the TM (1,0) cavity modes are excited simultaneously and mix each other. The mixture of these modes results in the generation of the circularly-polarized THz waves. We have also showed that the ellipticity of the THz wave dramatically changes with the applied voltage. The ellipticity reflects the relation between the amplitudes of the cavity modes that changes with the voltage. The way of the generation of the circularly-polarized THz waves and the control of their polarization discussed in this paper will widely extend the potential application of the IJJ-based THz emitter.

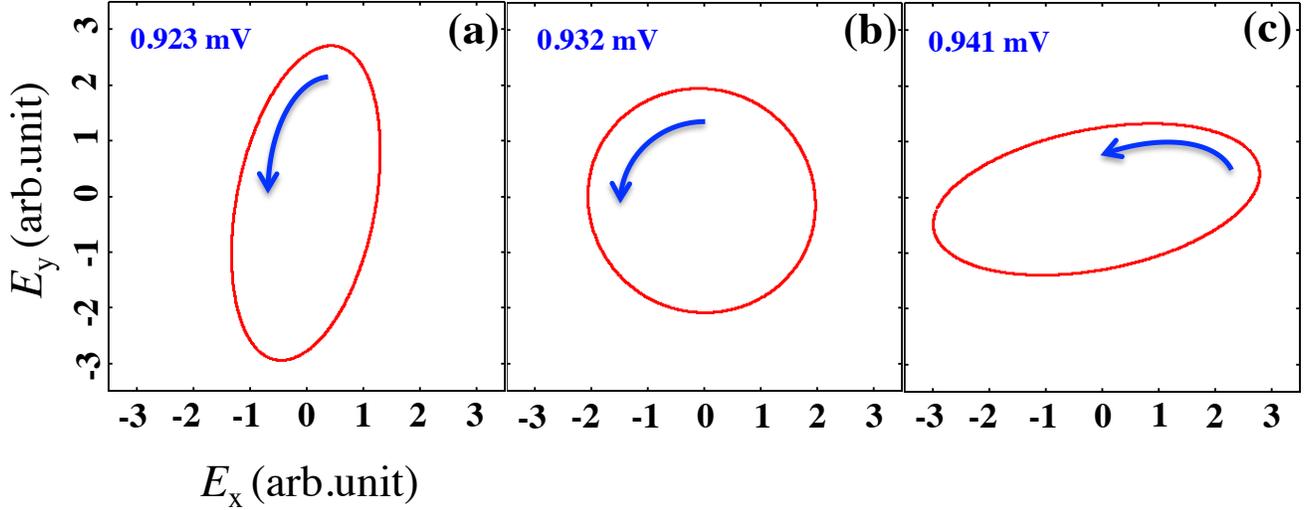

Fig. 3. Trajectories of the far electric fields along the $z$ axis at (a) 0.923 mV, (b) 0.931 mV, and (c) 0.941 mV.